\documentclass[aps,prl,twocolumn,groupedaddress]{revtex4}

\usepackage{amsmath}
\usepackage{amssymb}
\usepackage{hyperref}
\newcommand{\eqn}[1]{(\ref{#1})}

\begin{document}


\title{Non--Einstein source effects in massive gravity.}


\author{S.~Deser}
\email[]{deser@brandeis.edu}
\affiliation{Lauritsen Lab, Caltech, Pasadena CA 91125 and Physics Department, Brandeis University, Waltham, MA 02454, USA}

\author{A.~Waldron}
\email[]{wally@math.ucdavis.edu}
\affiliation{Department of Mathematics, University of California, Davis, CA 95616, USA}


\date{\today}

\begin{abstract}
We exhibit novel effects (absent in GR) of sources in massive gravity. First, we show that removing its ghost mode forces a field-current identity: The metric's trace is {\it locally} proportional to that of its  stress tensor; a point source implies a metric singularity enhanced  by the square of the graviton's range. Second, exterior solutions acquire spatial stress hair--their metric components depend on the interior  $T_{ij}$. Also, in contrast to  na\"ive
expectations, the Newtonian potential of a source is now determined by both  its  spatial stress and mass.  Our explicit results are obtained at linear, Fierz-Pauli, level, but qualitatively 
persist nonlinearly. 
\end{abstract}

\pacs{04.50.Kd, 04.30.Db,04.50.-h}

\maketitle

\section{Introduction}

Massive gravity (mGR) remains a growth industry (for an early catalog, see~\cite{mGR}), despite its numerous problems; for example, removing its ghost mode allows faster-than-light, acausal, propagation of the remaining, ``physical'' ones~\cite{DW}.
Here we exhibit some novel differences from GR in presence of matter, already at linear level~\footnote{These deviations are absent in the vector, Proca, analog of mGR. Consider  a static  source. The relevant equation, $\vec\nabla\cdot \vec E+m^2 A_0=0$ $\Rightarrow$ $ (\Delta -m^2)A_0 =\rho$ (the second equality holds because the erstwhile ``gauge fixing'', $\partial_\mu A^\mu=0$, is now a consequence of the Bianchi identity). In the present, static, case this means that $\vec \nabla\cdot\vec  A=0$, which excludes spherically symmetric $A_i$, leaving only the Yukawa equation for $A_0$; the electrostatic force is Yukawa, with no surprises. }. The first deviation, reminiscent of the old field-current identity (FCI)~\cite{LZ}, is that the metric's and stress tensor's traces are {\it locally} proportional; thus  a point source produces a naked metric singularity.  
A second one is the loss of the no-hair theorem: static, spherically symmetric, conserved, spatial, stresses contribute to the exterior metric (not just through the FCI), contrary to bald  GR~\cite{DL AJP}. The FCI effect is enhanced by an enormous factor $m^{-2}$,  and becomes singular in the massless limit. Further,  mGR's Newtonian potential  depends on the source's interior spatial stresses  as well as  its mass.

In Planck units, the linear, Fierz-Pauli (FP) field equations are
\begin{equation}\label{FP}
 {\cal G}_{\mu\nu}:= G^L_{\mu\nu}(h) -{\small \frac12}\,  m^2\big(h_{\mu\nu} - \eta_{\mu\nu}  h\big) = \frac12 \, T_{\mu\nu}\, ,
 \end{equation}
where $G^L_{\mu\nu}$ is the linearized Einstein tensor and the mass term is the unique ghost-free combination. 
We assume conservation of the source $T_{\mu\nu}$~\footnote{This universal assumption is, of course not strictly consistent for FP since matter, being coupled to $h_{\mu\nu}$, is no longer isolated.}.
The usual double-divergence of~\eqn{FP}  implies vanishing of the linearized scalar curvature---here the, non-gauge fixing, $\square h-\partial.\partial.h$, combination---which  means, upon tracing~\eqn{FP}, that
\begin{equation}
3m^2 h=  T\, .\label{h=T}
\end{equation}
This is our  FCI. It {\it locally}  equates  the traces of the metric's $h_{\mu\nu}$ and source's $T_{\mu\nu}$, and (except for null sources) is discontinuous in the $m\to0$ limit. A point mass thus induces a delta function singularity  $M \delta^3(r)$  in  $h:=h_{\mu}^\mu$. 
Note that even at a  finite (physical)  mass, there is an enormous enhancement of the metric by the square of mGR's range $\ell^2:= m^{-2}$.
Furthermore, this is not merely a linearized theory artifact, but is inherent 
to non-linear, 5 degrees of freedom (DoF), mGR extensions with algebraic mass terms~\cite{dRGT}: There, the left hand side of~\eqn{h=T} is supplemented by terms involving at most first metric derivative, contorsion, terms~\cite{DMZ,DW,DSW}\footnote{For one of the three allowed, 5 DoF, mass terms, a covariant scalar constraint is not known~\cite{DMZ}, but a non-covariant version of this result must hold.}, so that an  FCI persists in full~mGR. 

\section{Details}

The calculation involves the solution of the source-ful~FP equations~\eqn{FP},    using the standard massive spin~2 propagator,
\begin{equation}\label{prop}
\Delta_{\mu\nu \alpha\beta} \sim \frac{  P _{\mu(\alpha} P_{\beta)\nu}  - \frac13 P_{\mu\nu} P_{\alpha\beta}}{p^2+m^2}  \, ,
\end{equation}
defined in terms of the {\it on-shell} projectors
$$P_{\mu\nu}:=\eta_{\mu\nu}+\frac{p_{\mu}p_{\nu}}{m^2}\, .     $$
The essential difference between these projectors and those of massless spin 2 is that the latter involve, not $p_\mu p_\nu/m^2$, but $p_\mu p_\nu/p^2$: Because the massive one's trace has a $p^2/m^2$ numerator,  it can cancel the propagator's denominator, while the massless one just traces to unity. Using~\eqn{prop}, the metric produced by a conserved $T_{\mu\nu}$ is 
\begin{equation}
h_{\mu\nu} = \Delta_{\mu\nu\alpha\beta} T^{\alpha\beta} = \frac{T_{\mu\nu} - \frac13 P_{\mu\nu} T}{\square -m^2}   \Longrightarrow 
h = \frac{T}{3m^2}\, .
\end{equation}
The interaction $\int h_{\mu\nu} t^{\mu\nu}$ that this implies between two sources is {\it not} (quite) the usual
\begin{equation*}
I\sim \int \Big[t^{\mu\nu} \frac{1}{\square -m^2}\, T_{\mu\nu} -\frac13\,  t\,  \frac{1}{\square-m^2}\,  T \Big]\, ,
\end{equation*}
because reaching the latter required integrating the $p_\mu p_\nu$ term of $P_{\mu\nu}$ in~\eqn{prop} by parts, thereby implicitly losing  the FCI term: while indeed $\partial.t_\nu =\partial. T_\nu=0$, the trace part $p^2/m^2$ of $P_{\mu\nu}$ has been lost in the process! [In the massless case, there is no ambiguity since tracing $p_\mu p_\nu/p^2$ yields unity.]
This new contribution, $\delta I$, to the interaction is proportional to $t_\alpha^\alpha\,  h$, hence by~\eqn{h=T}, 
\begin{equation*}
\delta I \sim \int t\,  h\  = \  \frac1{3m^2} \int t\,  T\, .
\end{equation*}
This is a contact interaction between two (non-null) sources, so not really observable macroscopically,
but it might be seen in their s-wave scattering--which would be enormously magnified by the square of mGR's range.  To be sure, linearized theory misses any horizon-like features of the full nonlinear extension, so we cannot make realistic estimates of the effect. Note  that the full~$T_{\mu\nu}$'s trace is involved, so already here we see the presence of hair, {\it i.e.},   interior, spatial,~$T_{ij}$ components, even of spherically symmetric sources, contribute. [Since we are working at linear level, we can treat this source independently of any $T_{00}(r)$.]
Turning now to the hair effect itself, we must consider a {\it local}, spherically symmetric, static, conserved source $T_{ij}(r)$.
Hence, by conservation, this $T_{ij}$ must be the transverse projector of some scalar~\footnote{Absent spherical symmetry, $T_{ij}$ also admits transverse-traceless components.}. The
na\"ive ansatz $T_{ij}=\frac12  (\delta_{ij} - \partial_i \partial_j \Delta^{-1}) T(r)$ is disallowed because only its trace 
is local. Instead, locality can be restored by writing
\begin{equation}\label{nine}
T(r)= \Delta \tau(r)\Rightarrow T_{ii}=\Delta \tau\, ,
\end{equation}
for some local $\tau(r)$. Such sources are not physical in GR because they yield no external interactions (see below),
thereby upholding the traditional no-hair lore. As we shall see, the situation in mGR is rather different.

Inserting the source~\eqn{nine} into~\eqn{prop}, in terms of the Yukawa potential $Y:= (\Delta -m^2)^{-1}$, one finds 
\begin{eqnarray}\label{ten}
h_{ij}  &=&  \quad \int d^3 x \, Y\Big[T_{ij} -\frac13 \Big(\delta_{ij}+\frac{p_i p_j}{m^2} \Big)T\Big]\, ,  \nonumber\\[3mm]
h_{00}&=& \frac13\int d^3 x\; Y\, T\, ,
\label{ten}
\end{eqnarray}   
(whose trace $h=-h_{00}+h_{ii}$ indeed obeys the~FCI~\eqn{h=T}).
Hence, 
mGR's Newtonian potential $h_{00}$  depends on spatial stresses. More precisely, when the interior $\tau$ and exterior $t_{00}$ do not overlap, their hairy Newtonian interaction is $$I\sim  \int t_{00} Y \Delta \tau=\int t_{00}\big[ 1 + m^2  \,  Y\big]\, \tau
=m^2  \int t_{00} \, Y\,  \tau\, .$$
While non-vanishing, this new interaction is suppressed by the product $(mR)^2$ where $R$ is the radius of the source~\footnote{In contrast, 
a slow GR probe (with only~$t_{00}$) interacts with the interior  (see~\cite{DL AJP} or Eq.~\eqn{ten}) according to
$I\sim \int t_{00}\,  C\, T$, where $C$ is the Coulomb potential. By virtue of  Eq.~\eqn{nine}, $T=\Delta \tau$ so
$I\sim\int t_{00}\,  C\,  \Delta \tau=\int t_{00} \, \tau$, which (as promised) vanishes when the exterior $t_{00}$ and interior $\tau$ do
not overlap.}. Similarly, the spatial stress interaction is $\int t^{ij} h_{ij}\sim  -\frac13 m^2  \int t_{ii} \, Y\,  \tau$.
The third possible static conserved source  
$T_{0i} =\varepsilon_{ijk}\partial_j J_k$ generates $h_{0i}$, leading to a spin-spin interaction, since
 $T_{0i}(r) \sim \epsilon_{ijk} \partial_j J_k \delta^3 (r)$ is both local and conserved. 
  The resulting   interactions are identical to the textbook Maxwell magnetic dipole-dipole's, including a $\vec j \cdot\vec J$ contact term.
In mGR there are forces beyond Maxwellian ones, since the derivatives in the momentum densities
now act on $Y$, rather than just $C$~\cite{BT}; there is also a novel Yukawa term~$\sim  m^2\! \int \!j\,  Y J$.

\section{Conclusion}
Using the linearized, FP, approximation to mGR, we  found novel properties of mGR's  sourceful solutions that
deviate  from  GR-based exterior solution  expectations~\footnote{The 
  GR interaction formula $I=\int  [t_{\mu\nu} \, C\,  T^{\mu\nu}-\frac12{t}_\mu^\mu\, C\,  T_\nu^\nu]$ ostensibly involves all components $T_{\mu\nu}$,
  which seems to conflict with the 
   absence of any long-range $T_{ij}$ effects and the ``no hair'' GR mantra that spherical exterior metrics only ``see'' $T_{00}$.
   The resolution is as follows:
First,  there is no $T_{0i}$ hair because there exist no spherically symmetric transverse vectors/dipoles.
 Second,  the interaction~$I$ comes from the gravitational field, while $T_{\mu\nu}$ must separately obey locality and conservation constraints. Locality of $T_{\mu\nu} (x)$ is a physical requirement--its local values matter when they become sources of a field, unlike in special relativity, where only their space integrals, the Lorentz generators, or the total charge for currents $j_\mu(x)$, are physical (conservation is of course also required by the GR field's Bianchi identities). 
 Locality and conservation of a time-independent source,  such as a local tensor $T_{ij}\sim \delta_{ij} \delta^3(r)$, requires adding a long-range tail $\sim  \partial_i\partial_j (1/r)$. Reinstating locality entails multiplying this by a $\Delta$ factor. But then the resulting, ultralocal, $T_{ij} \sim (\delta_{ij} \Delta -\partial_i \partial_j)\delta^3 (r)$ precludes long-range interactions--the extra $\Delta$ removes the Coulomb denominator (see~\cite{DL AJP}). An alternate explanation of GR's lack of $T_{ij}$ effects  is that  its  counterpart of~\eqn{ten}, namely (up to a gauge) $h_{ij}= C (T_{ij} -\frac12 \delta_{ij} T)= -\frac12 C \partial_i\partial_j \tau(r)$. But this is precisely a gauge, hence irrelevant. The above considerations do not contradict the usual application of $I$ to light bending: Its Maxwell fields are time dependent, hence involve spacetime-, rather than just space- conservation.}. While these  were obtained explicitly at linear level, they should persist qualitatively in full mGR. Specifically, we found: 
 (i) The scalar constraint is proportional to  $m^{-2}$ times the trace of the source 
$T_{\mu\nu}$--in particular,
the metric's trace acquires a delta-function singularity for a point source. 
(ii) Black mGR  holes have  hair--indeed their (spherically symmetric, static, conserved) spatial stresses affect the entire exterior metric~$h_{ij}$, not just its trace in the FCI. The contact terms are enhanced by the square of mGR's range and are singular in the massless limit. 
(iii) The Newtonian potential now also depends--though it is highly damped--on the interior stresses. 

\begin{acknowledgments}
We thank Bayram Tekin for useful discussions and A.W.  thanks the Perimeter Institute for hospitality while this research was completed.
S.D. was supported in part by NSF Grant No. PHY- 1064302 and DOE Grant No. DE-FG02-164 92ER40701. 
\end{acknowledgments}


\end{document}